\documentclass{article}

\usepackage[nonatbib,final]{neurips_2020}

\usepackage{graphicx}
\usepackage[utf8]{inputenc} 
\usepackage[T1]{fontenc}    
\usepackage{hyperref}       
\usepackage{url}            
\usepackage{booktabs}       
\usepackage{amsfonts}       
\usepackage{nicefrac}       
\usepackage{microtype}      

\title{Identifying Opportunities for Skillful Weather Prediction with Interpretable Neural Networks}

\author{%
  Elizabeth A. Barnes\thanks{http://barnes.atmos.colostate.edu} \\
  Department of Atmospheric Science\\
  Colorado State University\\
  Fort Collins, CO 80526 \\
  \texttt{eabarnes@rams.colostate.edu} \\
  \And
  Kirsten Mayer \\
  Dept. of Atmospheric Science\\
  Colorado State University\\
  Fort Collins, CO 80526 \\
  \texttt{kjmayer@rams.colostate.edu} \\
  \And
  Benjamin Toms \\
  Intersphere, Inc.\\
  \texttt{ben.toms@intersphere.earth} \\
  \And
  Zane Martin \\
  Dept. of Atmospheric Science\\
  Colorado State University\\
  Fort Collins, CO 80526 \\
  \texttt{zkmartin@rams.colostate.edu} \\
  \And
  Emily Gordon \\
  Dept. of Atmospheric Science\\
  Colorado State University\\
  Fort Collins, CO 80526 \\
  \texttt{emily.gordon@rams.colostate.edu} \\
}

\begin{document}

\maketitle

\begin{abstract} 
The atmosphere is chaotic. This fundamental property of the climate system makes forecasting weather incredibly challenging: it's impossible to expect weather models to ever provide perfect predictions of the Earth system beyond timescales of approximately 2 weeks. Instead, atmospheric scientists look for specific states of the climate system that lead to more predictable behaviour than others. Here, we demonstrate how neural networks can be used, not only to leverage these states to make skillful predictions, but moreover to identify the climatic conditions that lead to enhanced predictability. Furthermore, we employ a neural network interpretability method called ``layer-wise relevance propagation'' to create heatmaps of the regions in the input most relevant for a network's output. For Earth scientists, these relevant regions for the neural network's prediction are by far the most important product of our study: they provide scientific insight into the physical mechanisms that lead to enhanced weather predictability. While we demonstrate our approach for the atmospheric science domain, this methodology is applicable to a large range of geoscientific problems.

\end{abstract}

\section{Introduction and Motivation}
\label{intro}

It is perhaps no surprise that the founder of chaos theory, Ed Lorenz, was trained as a meteorologist and studied predictability of weather and climate. Since Lorenz first demonstrated the mathematical idea of chaos \cite{Lorenz1963-kt} -- the concept that tiny differences in the initial condition of a complex system can lead to vastly different future states -- Earth scientists have wrestled with the consequences for weather forecasting. In particular, on ``subseasonal'' timescales ($\sim$ two to five weeks) and longer, the chaotic nature of the atmosphere is fully realized, and skillful predictions are very difficult. However, the societal need for subseasonal and seasonal forecasts is acute: providing \textit{any} useful climate information several weeks in advance may have enormous benefits \cite{National_Academies_of_Sciences2016-tj}.

To break the deadlock on this problem, Earth scientists have come to understand that while skillful prediction \textit{in general} is mathematically intractable on subseasonal-to-seasonal timescales, certain atmospheric processes or configurations of the Earth system can be leveraged to make accurate subseasonal predictions. While these states are not always present at a given moment in time, when they are, scientists have the \textit{opportunity} to use them to make skillful predictions: we call these instances \textbf{forecasts of opportunity.}
As scientists studying this problem, our goal is two-fold: (1) we must leverage the predictability of forecasts of opportunity to fully extract useful and accurate predictions, and (2) we must work to identify \textit{when} forecasts of opportunity exist to begin with.
While the use-case presented here is heavily based on an example from \cite{Toms2020-lj}, multiple domain-specific applications led by the co-authors are in preparation or under review \cite{Mayer2020-s2s, Toms2020-dec}.

\section{Data}
\label{data}
Our network takes as input anomalous monthly sea-surface temperatures (SSTs) across the globe from the Cobe V2 monthly SST anomaly data set \cite{Hirahara2014-ts}, which we linearly regrid to a 4 degree by 4 degree grid spacing. The original 1 degree by 1 degree version of the Cobe V2 SST data used here can be accessed via NOAA ESRL (https://www.esrl.noaa.gov/psd/data/ gridded/data.cobe2.html). Each gridded map contains 45 latitudinal grid boxes and 90 longitudinal grid boxes, flattened prior to input into the network as a vector of length 4050. Points over land are included, but are assigned a value of zero. As output, we use gridded monthly surface air temperature anomalies from the Berkeley Earth Surface Temperatures (BEST) \cite{Rohde2013-jx} data set. This data can be accessed at http://berkeleyearth.org/data/. For our use-case here, we use only the air temperature time series at the grid point closest to Seattle, Washington (50$^{o}$N, 240$^{o}$E). 

\section{Methods}
\label{methods}

\subsection{Network architecture and training}
\begin{figure}
  \centering
  \includegraphics[width=4.0in]{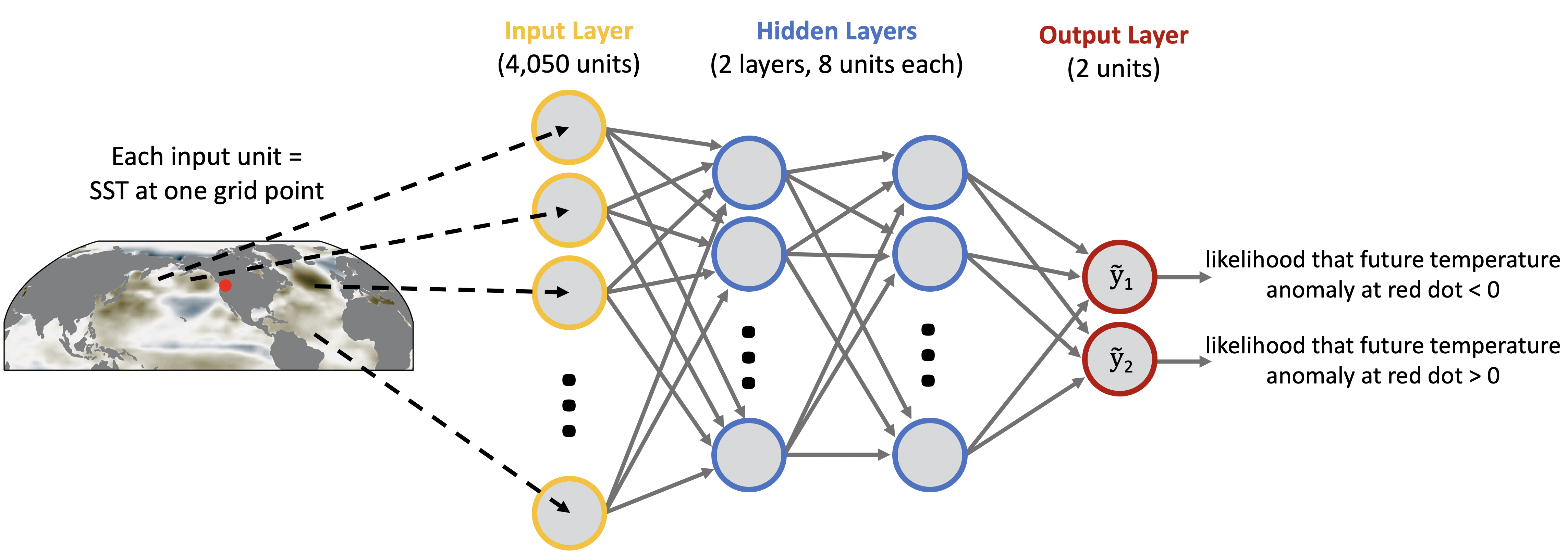}
  \caption{Neural network setup and architecture for predicting the average sign of the temperature anomalies over Seattle based on sea-surface temperatures (SSTs) 3 months prior.}
  \label{arch}
\end{figure}

We set the network up as a standard binary classification problem. The basic fully-connected, feed-forward network architecture is shown in Fig.~\ref{arch}. The input into the network is the flattened vector of SST from one month. The network then contains two hidden layers of 8 nodes each. Each hidden layer is followed by the Rectified Linear Unit (ReLu) activation function. The final output layer consists of two nodes representing whether anomalous air temperatures at our chosen location (e.g. near Seattle) is above and below zero. The output layer includes a soft-max layer to convert the output into likelihoods that sum to 1. The predicted class is computed as the class with the largest likelihood, or ``confidence'': our work will make use of these likelihood values when quantifying forecasts of opportunity.

The network is trained using the negative log likelihood loss using gradient descent with the Nesterov accelerated stochastic gradient descent optimizer and Nesterov momentum parameter of 0.9. The learning rate starts at 0.01 and is decreased to 0.005 after 50 epochs and trained for an additional 250 epochs. Given the large input size and substantial spatial correlation of global SSTs, we include L$_2$ regularization (i.e. ridge regularization) to spread the weights across the inputs and reduce overfitting. We set the L$_2$ parameter to 10.0 between the input layer and first hidden layer, and set it to 0.01 for all other layers. The training set spans all predictions within the year range of 1950-2005 (672 samples), and the testing set spans all predictions from 2006-2019 (168 samples). Input SSTs and output air temperatures are standardized by subtracting their respective 1980-2009 mean values at each grid point and linearly detrending each grid point to account to first order for anthropogenic climate warming. The range of 1980-2009 is a common climate baseline. 

For timescales beyond $\sim$ two weeks, it is standard to predict weather in an aggregate sense (i.e. averaged over a given time period) to capture lower frequency temperature fluctuations. Thus, we task the network with classifying the average temperature anomaly over Seattle 3 months in the future, which we refer to as a ``3 month forecast" henceforth. Recall the label associated with each sample is binary, predicting if the average temperature anomaly will be positive (assigned a label of 1) or negative (a label of 0).

\subsection{Layer-wise relevance propagation (LRP)}
Layer-wise relevance propagation (LRP) is a neural network intepretability method that produces a heatmap of the most relevant parts of the input for a given prediction. The papers which introduced LRP describe the full method in detail \cite{Bach2015-hi,Montavon2017-jg}. \cite{Toms2020-lj} illustrates the method from the perspective of a domain scientist, and \cite{Barnes2020-vt} demonstrates its use for climate change applications. 

In short, LRP takes the network output for a specific class (prior to the softmax layer), termed ``relevance'', and propagates the value backward through the network according to a list of propagation rules. Upon reaching the input layer, the relevance is split across the input units (in our use-case, the SST vector), where larger values imply that the particular unit is more relevant for that specific network prediction. Unlike backward optimization, LRP produces a separate heatmap for each prediction (i.e. sample), and thus, one can explore the unique decision-making process for each sample. One main use of LRP within the field of computer science appears to be for debugging a particular network setup or gaining trust in the network's output \cite{Lapuschkin2019-ni}. Yet we believe LRP, applied to particular questions in other scientific domains, stands to be transformative for advancing and discovering new science, as suggested by \cite{Toms2020-lj,Barnes2020-vt,Bohle2019-uw}. As one example, here we demonstrate how LRP can provide \textit{scientific insight} into the physical mechanisms that lead to enhanced predictability of surface air temperature on subseasonal-to-seasonal timescales.

\section{Results}
\begin{figure}
  \centering
  \includegraphics[width=2.75in]{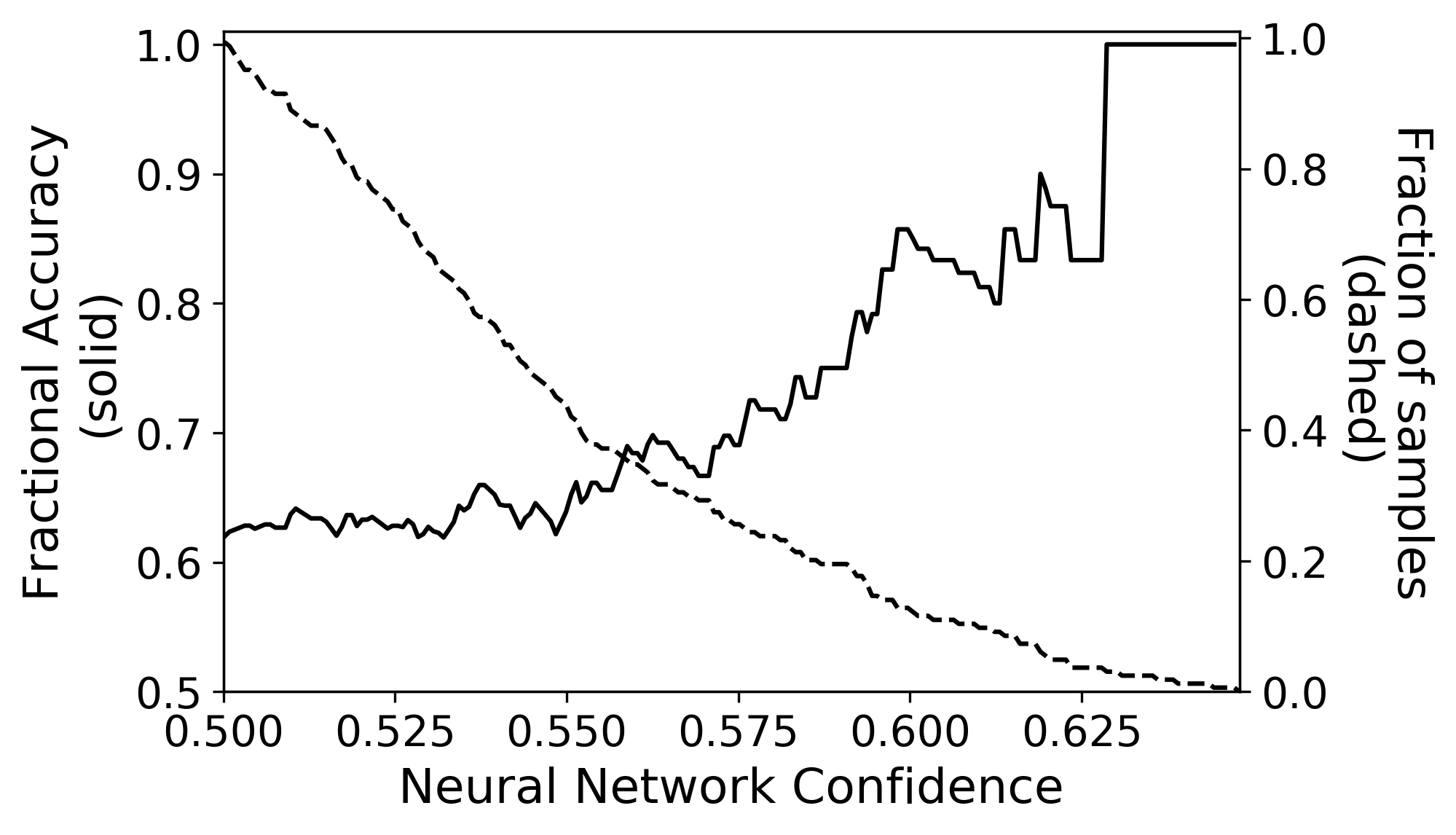}
  \caption{(solid curve) Accuracy of the neural network on the testing data as a function of the network confidence threshold. (dashed curve) Fraction of samples with network confidence above the threshold, and thus, the fraction of samples used to compute the accuracy.} 
  \label{foo}
\end{figure}
Predictions of the testing samples show an accuracy of 62\% (training accuracy of 65\%). 
As discussed above, we expect some climate states to lead to more predictable behaviour than others, and we can use the ANN to identify these states via its predicted confidence/likelihood. Fig. \ref{foo} shows the accuracy across the testing samples as a function of the confidence threshold used: as the network confidence increases, so does the accuracy. For example, the top 50\% most confident predictions have an accuracy of 63\% while the top 10\% most confident predictions have an accuracy of 82\%. For the more confident samples performance is much improved compared to 62\% accuracy across all testing samples. That is, the ANN's predicted confidence acts as a metric for identifying climate states that are generally more predictable. This sets the stage for our ability to then identify the physical drivers and mechanisms of these forecasts of opportunity. Of course, the number of samples decreases as the confidence threshold is increased, suggesting that as we increase our confidence threshold, the number of opportunities for making a forecast decreases. 

\begin{figure}
  \centering
  \includegraphics[width=4.5in]{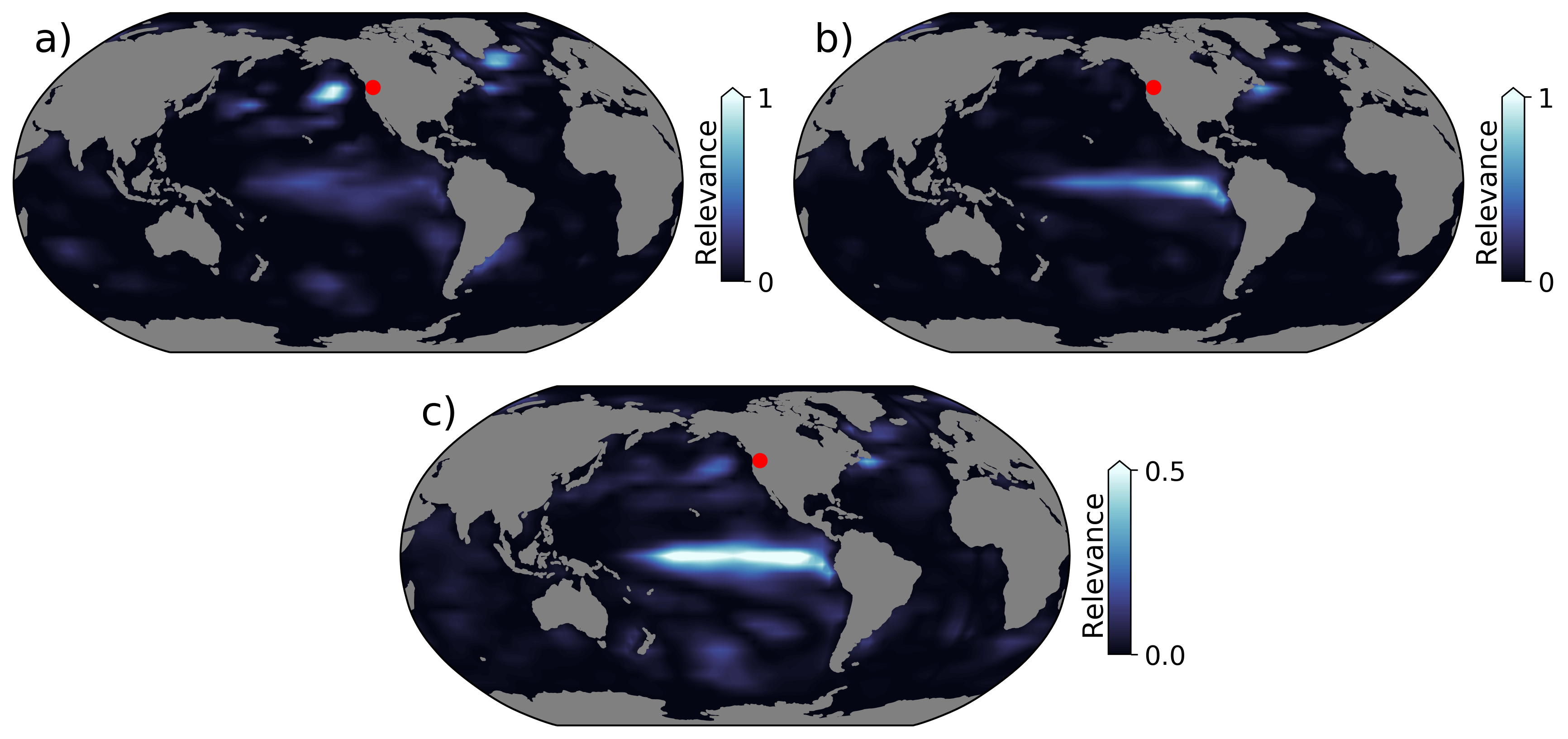}
  \caption{(a)-(b) Heatmaps of SST from LRP for two example accurate predictions. (c) Average of LRP heatmaps for accurate predictions in the top 10\% most confident predictions across training and testing samples (N=63).}
  \label{lrp}
\end{figure}

While identifying forecasts of opportunity for temperature predictions 3 months into the future is useful in its own right, scientifically from our perspective it is more interesting to explore the SST patterns that correspond to these predictions. To do so, we use LRP to generate heatmaps of the most relevant regions of SST for the ANN's prediction. Fig. \ref{lrp}a,b show relevance heatmaps for two samples that produced accurate predictions. The lightest colors show the regions most relevant for the ANN's accurate prediction. We observe high relevance in a small region of the midlatitude eastern Pacific in (a) and over the tropical eastern Pacific in (b) (indicative of the El Nino Southern Oscillation, or ENSO, a well known driver of midlatitude temperature variability). Note too that, because each LRP heatmap is computed for each input sample separately, differences between Fig. \ref{lrp}a,b demonstrate the network refocusing its attention depending on the specific sample.

While Fig. \ref{lrp}a,b show individual samples, Fig. \ref{lrp}c displays the average over many the LRP heatmaps: here we average over the top 10\% most confident predictions that were accurate in training and testing. That is, we take the mean of those predictions that we might categorized as confident \textit{forecasts of opportunity}. The average LRP map shows that the ENSO region emerges as a dominant driver of predictability, however, individual heatmaps (such as Fig. \ref{lrp}a) indicate that the network uses other sources as well. Furthermore, these high-relevance regions align well with areas identified in recent work on the physical mechanisms for temperature prediction on these timescales \cite{Toms2020-lj,Capotondi2019-hb}.

\section{Concluding Thoughts}
We demonstrated how interpretable neural networks can be used to filter through the noise of a chaotic climate system and identify periods of enhanced predictability, or ``forecasts of opportunity''. Physical insight into the mechanisms behind this enhanced predictability was then gained (e.g. the ENSO region) by using layerwise relevance propagation to identify the regions of the input (maps of sea-surface temperatures) that were most relevant for the network's accurate prediction. We note that this methodology alone does not account for causation, as the layerwise relevance propagation heatmaps merely provide the regions in the input that provide predictability. With that said, these heatmaps can then be interpreted by a domain scientist to further investigate the dynamical mechanisms that may tie the input to the prediction of the output in a causal sense. This could include investigating the time evolution of the sample fields that are identified as successful forecasts of opportunity.

Next steps for this work include exploring additional cost functions and network setups to further encourage the network to focus on climate states that lead to enhanced predictability, rather than minimizing the cost function over all samples. Further, future work will focus on continuing to understand the physical mechanisms in the ocean and atmosphere of the sources of predictability identified in this study, in Seattle but also at other points around the globe. While we demonstrate our approach for the atmospheric science domain, this methodology is applicable to a large range of geoscientific problems.

\section*{Broader Impact}
The broader impacts of this work within atmospheric science include improving weather and climate predictions across seasonal to decadal time scales, while more broadly in the geosciences the ``forecast of opportunity'' framework is applicable across a wide range applications. Ultimately, accurate and timely forecasts can save lives and property, while poor forecasts put both in jeopardy. Investors, farmers, and energy managers all stand to benefit from skillful and useful subseasonal forecasts, while inaccurate or over-confident forecasts could lead to poor allocation of resources, financial loss, and food shortages. Thus, extensive testing of any forecast product based on this work is necessary and essential prior to it being used for any policy and decision-making. Additionally, as the climate changes with anthropogenic warming and we enter new and less well-understood climate regimes, models should be constantly reexamined and re-calibrated to the extent possible to ensure they stay relevant and accurate.

\begin{ack}
This work is supported by funding from the National Science Foundation under grants AGS-1749261, 2019758, 006784, 2020305, the Department of Energy under grant DE-FG02-97ER25308, and Fulbright New Zealand.

\end{ack}

\bibliographystyle{ieeetr}
\bibliography{neurips2020Barnes.bib}

\begin{thebibliography}{10}

\bibitem{Lorenz1963-kt}
E.~N. Lorenz, ``{Deterministic Nonperiodic Flow},'' {\em Journal of the
  Atmospheric Sciences}, vol.~20, pp.~130--141, Mar. 1963.

\bibitem{National_Academies_of_Sciences2016-tj}
{National Academies of Sciences} and {Medicine}, {\em {Next Generation Earth
  System Prediction: Strategies for Subseasonal to Seasonal Forecasts}}.
\newblock Washington, DC: The National Academies Press, 2016.

\bibitem{Toms2020-lj}
B.~A. Toms, E.~A. Barnes, and I.~Ebert‐Uphoff, ``{Physically Interpretable
  Neural Networks for the Geosciences: Applications to Earth System
  Variability},'' {\em Journal of Advances in Modeling Earth Systems}, June
  2020.

\bibitem{Mayer2020-s2s}
K.~Mayer and E.~Barnes, ``{{Subseasonal Forecasts of Opportunity Identified by
  an Interpretable Neural Network}},'' {\em Geophysical Research Letters},
  under review.

\bibitem{Toms2020-dec}
B.~Toms, E.~Barnes, and J.~Hurrell, ``{{Assessing Decadal Predictability in an
  Earth-System Model Using Interpretable Neural Networks}},'' {\em Geophysical
  Research Letters}, under review.

\bibitem{Hirahara2014-ts}
S.~Hirahara, M.~Ishii, and Y.~Fukuda, ``{Centennial-Scale Sea Surface
  Temperature Analysis and Its Uncertainty},'' {\em Journal of climate},
  vol.~27, pp.~57--75, Jan. 2014.

\bibitem{Rohde2013-jx}
R.~Rohde, R.~A. Muller, R.~Jacobsen, E.~Muller, S.~Perlmutter, A.~Rosenfeld,
  J.~Wurtele, D.~Groom, and C.~Wickham, ``{A New Estimate of the Average Earth
  Surface Land Temperature Spanning 1753 to 2011},'' {\em Geoinformatics \&
  Geostatistics: An Overview}, vol.~2013, 2013.

\bibitem{Bach2015-hi}
S.~Bach, A.~Binder, G.~Montavon, F.~Klauschen, K.-R. M{\"u}ller, and W.~Samek,
  ``{On Pixel-Wise Explanations for Non-Linear Classifier Decisions by
  Layer-Wise Relevance Propagation},'' {\em PloS one}, vol.~10, p.~e0130140,
  July 2015.

\bibitem{Montavon2017-jg}
G.~Montavon, S.~Lapuschkin, A.~Binder, W.~Samek, and K.-R. M{\"u}ller,
  ``{Explaining nonlinear classification decisions with deep Taylor
  decomposition},'' {\em Pattern recognition}, vol.~65, pp.~211--222, May 2017.

\bibitem{Barnes2020-vt}
E.~A. Barnes, B.~Toms, J.~W. Hurrell, I.~Ebert-Uphoff, C.~Anderson, and
  D.~Anderson, ``{Indicator patterns of forced change learned by an artificial
  neural network},'' {\em Journal of Advances in Modeling Earth Systems},
  vol.~n/a, p.~e2020MS002195, Aug. 2020.

\bibitem{Lapuschkin2019-ni}
S.~Lapuschkin, S.~W{\"a}ldchen, A.~Binder, G.~Montavon, W.~Samek, and K.-R.
  M{\"u}ller, ``{Unmasking Clever Hans predictors and assessing what machines
  really learn},'' {\em Nature communications}, vol.~10, p.~1096, Mar. 2019.

\bibitem{Bohle2019-uw}
M.~B{\"o}hle, F.~Eitel, M.~Weygandt, and K.~Ritter, ``{Layer-Wise Relevance
  Propagation for Explaining Deep Neural Network Decisions in MRI-Based
  Alzheimer's Disease Classification},'' {\em Frontiers in aging neuroscience},
  vol.~11, p.~194, July 2019.

\bibitem{Capotondi2019-hb}
A.~Capotondi, P.~D. Sardeshmukh, E.~Di~Lorenzo, A.~C. Subramanian, and A.~J.
  Miller, ``{Predictability of US West Coast Ocean Temperatures is not solely
  due to ENSO},'' {\em Scientific reports}, vol.~9, p.~10993, July 2019.

\end{thebibliography}

\end{document}